%% file: main.tex
\documentclass[a4paper,11pt]{article}
\usepackage{pos}
\usepackage{array}
\usepackage{booktabs}
\usepackage[sicmds,freestanding]{hepunits}
\usepackage{tikz}
\usetikzlibrary{arrows.meta, positioning, shapes.geometric, fit,positioning,decorations.markings}
\usetikzlibrary{calc}
\usepackage{subcaption}
\usepackage{makecell}

\title{Hardware-efficient neural networks for FPGA-based radio triggering of extensive air showers}
\ShortTitle{FPGA-based neural networks for radio triggering of extensive air showers}

\author*[a,b]{Vesselin Dimitrov}
\author[a]{Alperen Aksoy}
\author[a]{Ilja Bekman}
\author[b]{Markus Cristinziani}
\author[b]{Eric-Teunis de Boone}
\author[b]{Qader Dorosti}
\author[a]{Chimezie Eguzo}
\author[c]{Stefan Heidbrink}
\author[a,d]{Stefan van Waasen}
\author[a]{Andre Zambanini}

\affiliation[a]{Peter Gr\"unberg Institute -- Integrated Computing Architectures (ICA | PGI-4), Forschungszentrum J\"ulich GmbH,\\
Wilhelm-Johnen-Straße, J\"ulich, Germany}
\affiliation[b]{Center for Particle Physics Siegen, Department Physik, Universit\"at Siegen,\\ 
Walter-Flex Straße 3, Siegen, Germany}
\affiliation[c]{Elektronikentwicklungslabor, Department Physik, Universit\"at Siegen,\\
Walter-Flex Straße 3, Siegen, Germany}
\affiliation[d]{Faculty of Engineering, Communication Systems,  University of Duisburg-Essen,\\
Forsthausweg 2, Duisburg, Germany}

\emailAdd{dimitrov@hep.physik.uni-siegen.de}
\emailAdd{a.aksoy@fz-juelich.de}
\emailAdd{i.bekman@fz-juelich.de}
\emailAdd{markus.cristinziani@cern.ch}
\emailAdd{deboone@hep.physik.uni-siegen.de}
\emailAdd{Qader.Dorosti@uni-siegen.de}
\emailAdd{c.eguzo@fz-juelich.de}
\emailAdd{stefan.heidbrink@uni-siegen.de}
\emailAdd{s.van.waasen@fz-juelich.de}
\emailAdd{a.zambanini@fz-juelich.de}

\abstract{We present a hardware‑efficient hybrid trigger for FPGA‑based radio detection of extensive air showers. The hybrid design consists of a lightweight denoiser that cleans raw ADC traces and a compact classifier that operates on the denoised output, enabling robust near‑threshold pulse detection in high‑interference environments. Both neural networks are trained quantization‑aware. Signals are generated from detector‑folded CoREAS/CORSIKA simulations and embedded into measured noise to form a realistic benchmark. The trigger reaches an AUC of 0.992 while fitting comfortably within the resource budget of a Zynq‑7000 Z‑7020, with microsecond‑scale latency and sub‑watt power consumption. RTL validation confirms agreement between the fixed‑point hardware and the quantized software model, demonstrating that neural denoising combined with classification provides reliable, low‑cost radio triggering in noisy environments.}

\FullConference{11th International Workshop on Acoustic and Radio EeV Neutrino Detection Activities (ARENA2026)\\
8-11 June 2026\\
Karlsruhe, Germany\\}

\begin{document}
\maketitle
\flushbottom

\input{sections/01_introduction}

\input{sections/02_data_sets_and_signal_construction}
\input{sections/03_hybrid_trigger_method}
\input{sections/04_trigger_performance}
\input{sections/05_fpga_implementation}
\input{sections/06_discussion}
\input{sections/07_summary_and_outlook}

\acknowledgments
\input{sections/08_acknowledgments}

\bibliographystyle{JHEP}
\bibliography{sections/09_literature}

\end{document}

%% file: sections/01_introduction.tex
\section{Introduction}
\noindent
Ultra-high-energy cosmic rays (UHECRs) are particles originating from space and are observed using particle detectors, fluorescence telescopes, and radio antennas \cite{auger,TA_surface_detector,SCHRODER20171}. When a primary particle enters the atmosphere, it initiates an
extensive air shower (EAS) whose electromagnetic component emits radio waves through the
geomagnetic and charge-excess mechanisms \cite{geomEffect,AskEffect}. These radio signals encode key shower properties,
such as the energy and depth of the shower maximum \cite{Aab2016-fq,Abdul_Halim2024-ll}.

Autonomous radio triggering is attractive because radio antennas provide high duty cycle, broad
sky coverage, and sensitivity to the electromagnetic component \cite{HUEGE20161}. Inclined air showers especially benefit from radio‑only triggering, since their particle footprint at ground is strongly attenuated \cite{Aab2016-fq}.

The main challenge is the radio background. Transient anthropogenic bursts and
narrow-band RFI produce numerous signal-like fluctuations, making simple threshold triggers
ineffective near the detection threshold \cite{Schmidt2011-xy,Torres_Machado2013-my,KELLEY2013133}.

In previous work of our group, we have demonstrated that an AI-enhanced self-trigger implementable on an FPGA achieves higher accuracy than a threshold-based trigger \cite{Dorosti_2025}. A continuation of this work \cite{QD_hybrid} shows a significant reduction of resource usage.

This work introduces a hybrid trigger combining a lightweight convolutional denoiser with a
compact classifier, and serves as an update to our group's previous study \cite{QD_hybrid}. The denoiser aims to recover weak pulses at waveform level, while the classifier evaluates the cleaned trace.
Hyperparameter optimization (HPO), quantization-aware training (QAT), and High Granularity Quantization (HGQ) \cite{Sun2026-ce} aim to reduce the resource footprint while maintaining high performance.



%% file: sections/02_data_sets_and_signal_construction.tex
\section{Data sets and signal construction}

\subsection{Measured noise and simulated pulses}
Background traces were recorded on the University of Siegen campus using a butterfly antenna,
low‑noise amplifier, and oscilloscope. Signals were band‑limited to $30$–$80\,\mathrm{MHz}$ and
sampled at $250\,\mathrm{MHz}$. The environment contains strong transient and narrow‑band RFI,
providing a realistic and deliberately challenging noise benchmark \cite{Dorosti_2025}.

Air‑shower pulses are generated with CoREAS/CORSIKA using event‑by‑event geometries \cite{Huege2025-dm} from
Auger Engineering Radio Array (AERA) \cite{Fuchs2012AERA} data. The Pierre Auger \emph{Offline} framework \cite{Argiro2007-gv} provides
antenna and electronics folding. The amplitudes correspond to signed 12-bit integer ADC values.\\
The signal strength is kept very low with near‑threshold pulses, where the SNR, defined as the mean squared Hilbert‑envelope amplitude divided by the noise RMS squared, is centered around 3 and spans values from $\approx0.1$ to $\approx15$ \cite{QD_hybrid}.



\subsection{Training and validation}
Three classes are used in this work: \texttt{background} (noise‑only), \texttt{pure signal} (detector‑folded pulses), and \texttt{signal}
(pulse injected into noise). The denoiser and classifier are trained together, with the training input being \texttt{background} and \texttt{signal} traces. The denoiser targets are constant zeros for \texttt{background} traces and the corresponding \texttt{pure signal} trace for the \texttt{signal} input. The subsequent classifier takes the denoised output as an input.\\
A standalone classifier that works without a denoiser is also developed and trained on the raw \texttt{signal} and \texttt{background}.
Held‑out datasets provide validation
and large \texttt{background} samples for low‑FPR evaluation. These datasets are disjoint from the training data.
Table \ref{tab:dataset_splits} gives an overview of the datasets used and their purpose.

\begin{table}[t]
\centering
\small
\begin{tabular}{>{\raggedright\arraybackslash}p{0.19\textwidth}>{\raggedright\arraybackslash}p{0.25\textwidth}r >{\raggedright\arraybackslash}p{0.27\textwidth}}
\toprule
Dataset & Current use in this work & Traces & Composition \\
\midrule
Training Dataset & Full model training & \num{150000} & \num{50000} \texttt{background}, \num{50000} \texttt{pure signal}, \num{50000} \texttt{signal} \\
Training Dataset & HPO training & \num{30000} & \num{7500} \texttt{pure signal}, \num{7500} \texttt{signal}, \num{7500} zeros ,\num{7500} \texttt{background} \\
Training Dataset & HPO cross-validation & \num{30000} & \num{7500} \texttt{pure signal}, \num{7500} \texttt{signal}, \num{7500} zeros ,\num{7500} \texttt{background} \\
Validation Dataset 1 & Full model validation & \num{150000} & \num{50000} \texttt{background}, \num{50000} \texttt{pure signal}, \num{50000} \texttt{signal} \\

Validation Dataset 2 & False-positive estimate & \num{1000000} & \texttt{background} only \\
\bottomrule
\end{tabular}
\caption{Datasets used in the present study.}
\label{tab:dataset_splits}
\end{table}

%% file: sections/03_hybrid_trigger_method.tex
\section{Hybrid trigger method}

\subsection{Concept and models}
The trigger operates on $128$‑sample traces. A compact denoiser first maps the noisy waveform
to a cleaned estimate; a compact classifier then evaluates the cleaned trace and outputs a trigger
score.
The results are compared to a classical peak-envelope threshold trigger.
After developing the models in \texttt{Keras} \cite{chollet2015keras}, firmware is generated with \texttt{hls4ml} \cite{fastml_hls4ml}, followed by high-level synthesis with \texttt{Vitis HLS 2024.1} \cite{vitisHLS2023} and implementation with \texttt{Vivado 2024.1} \cite{vivado2024}.

\begin{figure}[!htbp]
\centering

\begin{minipage}[c]{0.48\linewidth}
\centering

\begin{tikzpicture}[
    box/.style={draw, thick, rounded corners, align=center,
                minimum width=1.0cm, minimum height=0.8cm,
                font=\scriptsize},
    line/.style={-latex, thick},
    skip/.style={dashed, thick},
    node distance=0.3cm
]

\node[box] (in) {Input\\128×1};
\node[box, right=0.3cm of in] (q) {Quantize};
\node[box, right=0.3cm of q] (e1) {Conv1D\\k=4, f=4};

\node[box, below left=0.3cm and -1 cm of e1] (e2) {Conv1D+ReLU\\k=2, f=4};
\node[box, left=0.3cm of e2] (e3) {Conv1D\\k=4, f=4};

\node[box, below=0.3cm of e3] (d1) {Conv1D\\k=2, f=4};
\node[box, right=0.3cm of d1] (d2) {Conv1D+ReLU\\k=4, f=4};

\node[box, below right=0.3cm and -1 cm of d2] (d3) {Conv1D\\k=4, f=4};
\node[box, left=0.3cm of d3] (res) {Conv1D\\k=1, f=1};
\node[box, left=0.3cm of res] (out) {Output\\128};

\draw[line] (in) -- (q) -- (e1) -- (e2) -- (e3)
            -- (d1) -- (d2) -- (d3)
            -- (res) -- (out);

\node[font=\bfseries, above right=0.3cm and -1.3 cm of q] {Denoiser};

\end{tikzpicture}

\end{minipage}
\hfill
\begin{minipage}[c]{0.48\linewidth}
\centering

\begin{tikzpicture}[
    box/.style={draw, thick, rounded corners, align=center,
                minimum width=1.0cm, minimum height=0.8cm,
                font=\scriptsize},
    line/.style={-latex, thick},
    node distance=0.3cm
]

\node[box] (c_in) {Input\\128×1};
\node[box, right=0.3cm of c_in] (c_q) {Quantize};

\node[box, below left=0.3cm and -2.4cm of c_q] (c1) {Conv1D+ReLU+\\MaxPool\\k=4, f=2};
\node[box, left=0.3cm of c1] (c2) {Conv1D+ReLU+\\MaxPool\\k=2, f=4};

\node[box, below=0.3cm of c2] (c3) {Conv1D+ReLU\\k=4, f=6};
\node[box, right=0.3cm of c3] (c4) {Conv1D+ReLU\\k=4, f=6};

\node[box, below left=0.3cm and -1.2cm of c4] (gap) {GAP};
\node[box, left=0.3cm of gap] (dense) {Dense(1)};

\draw[line] (c_in) -- (c_q) -- (c1) -- (c2) -- (c3)
            -- (c4) -- (gap) -- (dense);

\node[font=\bfseries, above right=0.3cm and -2.15cm of c_q] {Classifier};

\end{tikzpicture}

\end{minipage}

\caption{Architecture diagrams of the denoiser (left) and classifier (right). Kernel size is abbreviated by $k$, the number of filters by $f$. The classifier used in the hybrid model trained on denoised traces is the same as the standalone classifier trained on raw traces.}
\label{fig:architectures}
\end{figure}
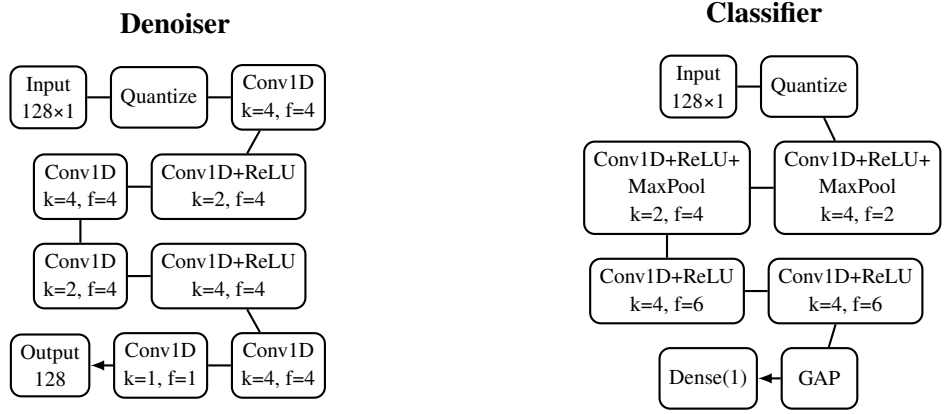

In the case of the hybrid model, both networks are 1D convolutional architectures. The denoiser has 607 trainable parameters, while the classifier has 598. The
architectures are shown in Fig.~\ref{fig:architectures}.

The two networks are trained jointly as a single composite model with two outputs; the denoised trace and classification logit. This ensures
that the denoiser and classifier share the same forward pass and receive coupled gradient updates
during backpropagation. Joint training prevents the denoiser from suppressing waveform features
that are informative for classification, and allows the classifier to adapt to the denoiser’s output distribution.

Training uses the Adam optimizer with a learning rate of $2.441\cdot10^{-4}$ and a batch size of $64$. The total loss is a weighted sum of the denoiser loss (mean squared error between denoised trace and the target) and the
classifier loss (binary cross-entropy). This setup encourages the denoiser to preserve
classification-relevant pulse characteristics while still reducing background fluctuations.

The standalone classifier utilizes Adam with a learning rate of $6.512\cdot10^{-3}$ and a batch size of 64. The architecture is the same as the classifier used in the hybrid composite model (see Fig. \ref{fig:architectures}).

\subsection{Hyperparameter optimization, quantization-aware training and High Granularity Quantization}
Hyperparameter optimization is performed on the composite model and the standalone classifier using HyperBand \cite{Li2018-hyperband}, covering both architectural and training hyperparameters as in \cite{QD_hybrid}. For the joint composite model, the relative weights of the denoiser and classifier losses are included in the optimization. The configuration achieving the lowest validation loss is selected as the final model.
Following the findings of \cite{qat_ptq}, which demonstrate that QAT preserves accuracy under aggressive quantization in resource‑limited edge hardware, all models in this work are trained quantization‑aware.
HGQ~\cite{Sun2026-ce} layers are used for all models, including during HPO.

The standalone classifier uses global bit widths of \texttt{ap\_fixed<30,15,RND,SAT>} with a $\beta-$value of $2.5\cdot10^{-5}$, describing the granularity of HGQ. The hybrid model's classifier uses \texttt{ap\_fixed<23,18,RND,SAT>} with $\beta=2.5\cdot10^{-6}$ and the denoiser \texttt{ap\_fixed<23,16,TRN,WRAP>} with $\beta=5\cdot10^{-6}$.


%% file: sections/04_trigger_performance.tex
\section{Trigger performance}

\subsection{Evaluation strategy}
Trigger performance is evaluated on the held-out dataset Validation~1 and a large
background-only sample (Validation~2). The performance metric is the receiver-operating-characteristic (ROC) curve; visualizing the true-positive rate (TPR) for \texttt{signal} traces vs. false-positive rate (FPR) for
\texttt{background} traces. The area under the curve (AUC) is also evaluated. Operating points are fixed at $\mathrm{FPR}_{\mathrm{nom}} = 10^{-2}$ and
$\mathrm{FPR}_{\mathrm{low}} = 10^{-4}$ and their corresponding $\mathrm{TPR}_{\mathrm{nom}}$ and $\mathrm{TPR}_{\mathrm{low}}$.
The classifier-only and hybrid model are compared to a classical peak-envelope threshold applied to the Hilbert amplitude.

\subsection{Neural branches}
The classifier-only branch reaches an AUC $=0.991$ and a $\mathrm{TPR}_{\mathrm{low}}\approx0.12$ and $\mathrm{TPR}_{\mathrm{nom}}\approx0.89$. \\
The hybrid trigger performs best, with a similar AUC $=0.991$ but higher $\mathrm{TPR}_{\mathrm{low}}\approx 0.36$ and $\mathrm{TPR}_{\mathrm{nom}}\approx0.96$. The classical baseline reaches an AUC $=0.629$ along with a $\mathrm{TPR}_{\mathrm{low}}\approx0.03$ and $\mathrm{TPR}_{\mathrm{nom}}\approx0.1$.

\begin{figure}[t]
    \centering
    \begin{subfigure}[t]{0.48\linewidth}
        \centering
        \includegraphics[width=\linewidth]{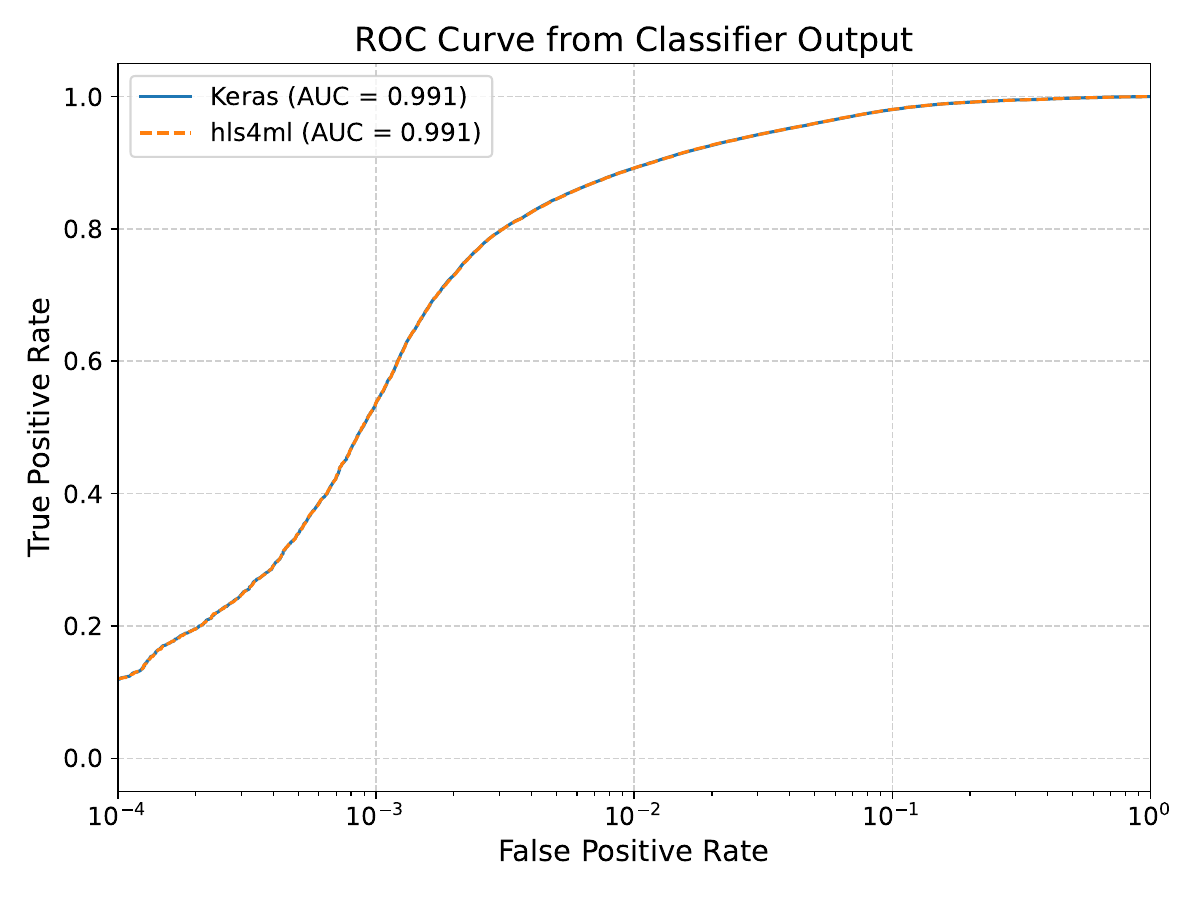}
        \caption{The ROC curve of the classifier.}
        \label{fig:roc_classifier}
    \end{subfigure}
    \hfill
    \begin{subfigure}[t]{0.48\linewidth}
        \centering
        \includegraphics[width=\linewidth]{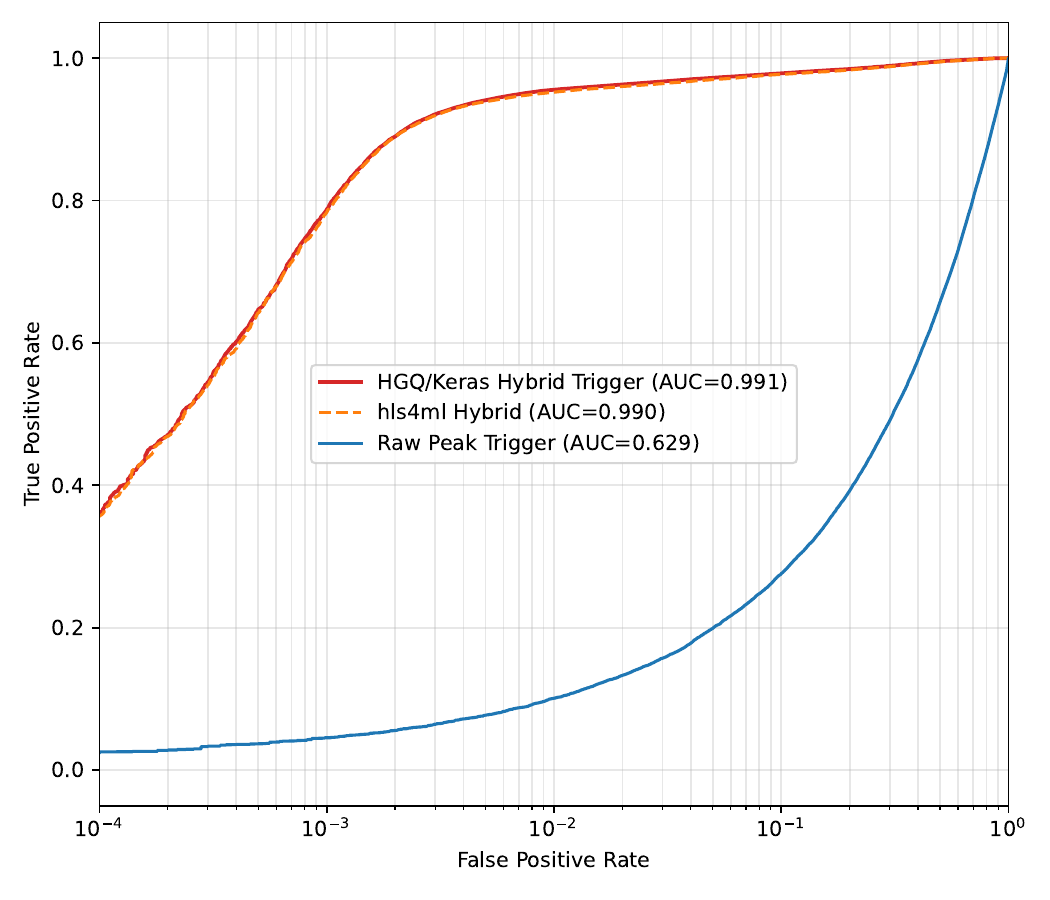}
        \caption{The ROC curve of the hybrid model.}
        \label{fig:roc_hybrid}
    \end{subfigure}
    \caption{The ROC curves of the neural networks. While the AUC of both models is the same, the hybrid model's signal efficiency improves for the low FPR regions, showcasing that the denoiser recovers trace features important for classification.}
\end{figure}

Figures \ref{fig:roc_classifier} and \ref{fig:roc_hybrid} show the ROC curve of the standalone classifier and of the hybrid trigger, which further improve the classification performance by reaching a higher AUC and signal efficiency for low FPR. The hybrid model in particular improves on the TPR for exceptionally low FPR compared to the standalone classifier, proving that the denoiser helps the classification performance by cleaning noisy traces.\\
What can be seen in both cases is that the \texttt{hls4ml} curve closely follows the \texttt{Keras} curve, showing that the models do not lose precision after imposing fixed point precision constraints on the layers.

%% file: sections/05_fpga_implementation.tex
\section{FPGA implementation}

The denoiser and classifier are exported with \texttt{hls4ml} and synthesized using \texttt{Vitis HLS 2024.1}.
The generated RTL blocks are integrated in \texttt{Vivado 2024.1} and validated with Validation Dataset 1. \\
All results were derived from the \texttt{Vivado} implementation of the hybrid trigger, which can be seen in Table \ref{tab:fpga_results}.
A Zynq-7000 Z-7020 is used as a conservative low-resource target FPGA. 
The implementation has an Initiation Interval equal to the model’s 12.7 µs inference time, yielding a throughput of approximately \num{78740} traces/s.

\begin{table}[t]
\centering
\small
\resizebox{\textwidth}{!}{%
\begin{tabular}{lccccccc}
\toprule
\raisebox{1ex}{Platform} &
\raisebox{1ex}{LUT} &
\raisebox{1ex}{FF} &
\raisebox{1ex}{BRAM} &
\raisebox{1ex}{DSP} &
\shortstack{Latency\\$[\mu\mathrm{s}]$} &
\shortstack{$F_{\max}$\\$[\mathrm{MHz}]$} &
\shortstack{dyn. Power\\$[\mathrm{W}]$} \\
\midrule
Zynq-7000 Z-7020 & \num{19042} & \num{31336} & \num{55.5} & \num{17} &
\num{12.7} & 161 & 0.502\\
\bottomrule
\end{tabular}
}
\caption{Post-implementation resource usage, latency, $F_{max}$ and dynamic power usage for the hybrid trigger on a Zynq-7000 Z-7020. LUT denotes lookup-table resources, FF denotes flip-flop resources, and BRAM denotes block random-access memory. The Z-7020 is used as a conservative low-resource reference.}
\label{tab:fpga_results}
\end{table}


RTL-level validation is performed using dedicated \texttt{cocotb} \cite{Gadde2024-rl} testbenches that stream
representative traces through the denoiser and feed the output to the classifier. The RTL outputs
are compared directly to the predictions of the \texttt{Keras} reference model.

For the denoiser, the testbench proves that the cleaned waveform matches the software
prediction within fixed-point tolerance.
For the classifier, the RTL logits are used to construct the ROC curve from
RTL outputs of the denoiser. The curve closely reproduces the software behaviour (RTL AUC of $\approx0.992$ vs. \texttt{Keras} AUC $\approx0.992$), confirming
that the fixed-point hardware simulation preserves the trigger performance.


%% file: sections/06_discussion.tex
\section{Discussion}

The hybrid trigger markedly improves the recovery of weak radio pulses at fixed background load.
The denoiser suppresses noise excursions while preserving the essential pulse morphology, which stabilizes the classifier and strengthens near‑threshold decisions. Residual distributions for both \texttt{signal} and \texttt{background} traces 
are centered around zero, indicating that the denoiser removes noise without distorting the underlying waveform.

Improved weak‑pulse efficiency reduces trigger‑level bias against inclined and radio‑faint showers. The compact FPGA implementation shows that these gains can be achieved within realistic station‑level resource budgets. 
The benchmark intentionally focuses on the weak‑signal regime, so absolute AUC values depend on this mixture. The denoised waveform is not intended as a precision reconstruction; it is best used for triggering and feature extraction rather than calorimetry.


%% file: sections/07_summary_and_outlook.tex
\section{Summary}

We presented a hybrid neural trigger for autonomous radio detection of extensive air showers.
The method combines a lightweight denoiser with a compact classifier, both trained
under strict hardware constraints using quantization-aware training and high-granularity fixed-point
quantization. Signals are constructed from detector-folded simulations and injected into measured high-interference noise, producing
a realistic near-threshold benchmark.

The hybrid trigger significantly improves weak-pulse efficiency at fixed false-positive rate,
outperforming both classical peak-envelope thresholds and classifier-only designs. The
denoiser enhances pulse visibility and stabilizes classifier performance, while the combined pipeline
achieves high AUC and the strongest TPR values. FPGA
implementation on a Zynq-7000 Z-7020 demonstrates that the trigger fits comfortably within
station-level resource budgets, with microsecond-scale latency and sub-watt power consumption.
End-to-end RTL validation confirms functional agreement with the quantized software model.

These results show that neural denoising and classification is a practical and effective strategy for radio-only
triggering in noisy environments, enabling improved sensitivity to inclined and radio-faint air
showers within realistic hardware constraints.

Compared to our previous work, a key advantage of the models developed here is that they operate
directly on raw ADC values without requiring input normalization. This further reduces resource
usage and simplifies deployment. Furthermore, the hybrid model presented here uses less resources at a similar performance.

%% file: sections/08_acknowledgments.tex
The authors express their sincere gratitude to the Electronics Laboratory of the Department Physik at Universität Siegen for their support with experimental electronics, access to laboratory infrastructure, and the hardware‑focused validation work essential to the measured‑noise and FPGA studies. We thank the Pierre Auger Collaboration for providing the simulation framework and detector‑response tools that enable the detector‑aware signal construction and the generation of the detector‑folded pulse library. We are also grateful to the \texttt{hls4ml} team, especially Enrico Lupi, for valuable discussions regarding the use and configuration of \texttt{hls4ml}. In addition, we acknowledge the developers and maintainers of the open‑source software that underpins this work, including \texttt{Python}, \texttt{NumPy}, \texttt{SciPy}, \texttt{Matplotlib}, \texttt{TensorFlow}/\texttt{Keras}, \texttt{hls4ml}, \texttt{cocotb}, \texttt{Vitis HLS}, and \texttt{Vivado}.